# Emergent Kondo scaling in iron-based superconductors $AFe_2As_2$ (A = K, Rb, Cs)

Y. P. Wu[1], D. Zhao[1], A. F. Wang[1], N. Z. Wang[1], Z. J. Xiang[1], X. G. Luo[1,2,4], T. Wu[1,2,4] and X. H. Chen[1,2,3,4]

1. Hefei National Laboratory for Physical Sciences at Microscale and Department of Physics, University of Science and Technology of China, Hefei, Anhui 230026, China

2. Key Laboratory of Strongly-coupled Quantum Matter Physics, Chinese Academy of Sciences, School of Physical Sciences, University of Science and Technology of China, Hefei, Anhui 230026, China

3. High Magnetic Field Laboratory, Chinese Academy of Sciences, Hefei, Anhui 230031, China

4. Collaborative Innovation Center of Advanced Microstructures, Nanjing University, Nanjing 210093, China

**Unconventional superconductivity from heavy fermion (HF) is always observed in *f*-electron systems, in which Kondo physics between localized *f*-electrons and itinerant electrons plays an essential role. Whether HF superconductivity could be achieved in other systems without *f* electrons, especially for *d*-electron systems, is still elusive. Here, we experimentally study the origin of *d*-electron HF behavior in iron-based superconductors (FeSCs) $AFe_2As_2$ (A = K, Rb, Cs). Nuclear magnetic resonance on $^{75}As$ reveals a universal coherent-incoherent crossover with a characteristic temperature T\*. Below T\*, a so-called "Knight shift anomaly" is first observed in FeSCs, which exhibits a scaling behavior similar to *f*-electron HF materials. Furthermore, the scaling rule also regulates the manifestation of magnetic fluctuation. These results undoubtedly support an emergent Kondo scenario for the *d*-electron HF behavior, which suggests the $AFe_2As_2$ (A = K, Rb, Cs) as the first material realization of *d*-electron HF superconductors.**

Superconductivity in heavy fermion (HF) materials is a conundrum in condensed matter physics. Conventional phonon-mediated pairing mechanism is failed in this case, suggesting an unconventional pairing mechanism similar to that in cuprates and organic superconductors (*1*). So far, only HF materials containing *f* electrons could be the hosts for unconventional superconductivity. Whether unconventional superconductivity could be achieved in HF materials without *f* electrons is still unknown. In *f*-electron HF materials, Kondo lattice has been widely accepted as the starting point for discussing the underlying physics (*2*). In two-fluid model of Kondo lattice (*3, 4, 5*), coherent state emerges below a characteristic temperature T\* as the localized *f*-electrons collectively reduce their entropy by hybridizing with the itinerant electrons to form a new state of matter, an itinerant heavy-electron Kondo liquid (KL) that displays scaling behavior. The emergent KL coexists with the hybridized spin liquid that describes the lattice of local moments whose magnitude has been reduced by hybridization. "Hybridization effectiveness" has been proposed as the organizing principle responsible for the emergence of low-temperature order in HF systems, including unconventional superconductivity (5). Although Kondo picture for HF has achieved a great success in *f*-electron materials, the fate of Kondo picture in HF materials without *f* electrons, especially for *d*-electron systems, is still controversial. Besides Kondo scenario (6), many alternative scenarios, such as lightly doped Mott insulator (7), geometrical frustration via antiferromagnetic interactions (8) and spin-orbital fluctuations (9), have been proposed to responsible for HF behavior in *d*-electron systems. In addition, searching for unconventional superconductivity in the existing *d*-electron HF materials is still not successful.

Recently, a remarkable mass enhancement was observed in heavily hole-doped Fe-based superconductors (FeSCs) $AFe_2As_2$ (A = K, Rb, Cs). As shown in Fig. 1, an apparent particle-hole asymmetry for effective mass (m*) appears in doped $BaFe_2As_2$ (10, 11, 12, 13). In heavily hole-doped $KFe_2As_2$, the value of Sommerfeld coefficient (γ) has reached to 98 mJ·mol·$K^2$ (10). Quantum oscillation and angle-resolved photoemission spectroscopy (APRES) experiments also support the mass enhancement (14, 15). Furthermore, the m* become even heavier by replacing $K^+$ by larger alkaline ions of $Rb^+$ (γ ~ 127 mJ·mol·$K^2$) (11) and $Cs^+$ (γ ~ 184 mJ·mol·$K^2$) (12). In FeSCs, the Hund's coupling is highly important for correlation effect instead of on-site Coulomb interaction (16), which has strong effect on band renormalization in FeSCs (17). However, recent theoretical calculation suggests that the remarkable mass enhancement in $AFe_2As_2$ (A = K, Rb, Cs) is beyond naive band renormalization (18). On the other hand, a coexistence of itinerant and local characters for 3*d* electrons has been widely recognized as an elementary feature in FeSCs (19), which hints a possible Kondo scenario for the mass enhancement in $AFe_2As_2$ (A = K, Rb, Cs). In fact, a coherent-incoherent crossover similar to *f*-electron HF has been already proposed for FeSCs (16), which is recalled by recent bulk magnetic susceptibility and thermal expansion experiments on $KFe_2As_2$ (20). However, there is still no consensus on Kondo picture experimentally. Here, we systematically perform $^{75}As$ nuclear magnetic resonance (NMR) experiment to study the hole-doped FeSCs $AFe_2As_2$ (A = K, Rb, Cs). Our results strongly support Kondo scenario for HF in $AFe_2As_2$ (A = K, Rb, Cs) and indicate that $AFe_2As_2$ (A = K, Rb, Cs) are *d*-electron HF superconductors under the frame of Kondo scenario.

In Fig.2, one of main findings in present work is a universal electronic crossover observed in both temperature-dependent Knight shift (K) and resistivity in $AFe_2As_2$ (A = K, Rb, Cs). In general, K = $K_{orb}$ + $K_s$, where $K_{orb}$ is the orbital contribution and $K_s$ is the spin contribution. $K_s$ is usually temperature-dependent, while $K_{orb}$ is always temperature-independent in FeSCs. $K_s$ is proportional to local spin susceptibility ($\chi_{loc}$) at the nearest-neighbour Fe atoms through transferred hyperfine coupling tensor ($K_s = A\chi_{loc}$). In a uniform paramagnetic state, since $\chi_{loc}$ is proportional to $\chi_{bulk}$, $K_s$ must be scaled with bulk magnetic susceptibility ($\chi_{bulk}$). Therefore, the temperature-dependent Knight shift could be expressed as $K_s(T) = A\chi_{bulk}(T) + K_{orb}$. As shown in Fig.2, the temperature-dependent part of Knight shift above T* can be fitted by a Curie-Weiss formula for all samples, suggesting a local-moments dominated behavior. Similar behavior has also been confirmed in bulk magnetic susceptibility measurements (See supplementary information). As temperature decreases below T*, Knight shift undergoes a crossover to deviate from the high-temperature Curie-Weiss behavior. There is a roughly 10% decreasing in the values of Knight shift for all samples. Finally, the temperature-dependent Knight shift becomes saturated at low temperature. The T* for AFe$_2$As$_2$ (A = K, Rb, Cs) shows a continuous suppression with replacing K$^+$ (T* ~ 165±25 K) by larger alkaline ions of Rb$^+$ (T* ~ 125±20 K) and Cs$^+$ (T* ~ 85±15 K). Although the broad linewidth made the crossover behavior indistinguishable in Knight shift in earlier experiments on KFe$_2$As$_2$ (*21*), bulk magnetic susceptibility and thermal expansion experiments showed a similar crossover behavior in KFe$_2$As$_2$ (*20*), consistent with the present NMR results. Surprisingly, the T* (~ 100 K) determined by the maximum of $\chi_{bulk}$ is much lower than that determined from Knight shift measurement (T* ~ 165±25 K). Besides KFe$_2$As$_2$, similar discrepancy of T* is also observed for RbFe$_2$As$_2$ and CsFe$_2$As$_2$. Impurity effect might be a reason for this discrepancy. In fact, a clear Curie-Weiss tail due to impurity effect is observed in the bulk magnetic susceptibility (See supplementary information), while it is absent in the Knight shift. A possible explanation for this difference is that NMR is a local probe for spin susceptibility and trivial impurities do not affect the Knight shift measured by $^{75}$As nuclei away from impurities. However, even if we try to subtract the Curie-Weiss tail from $\chi_{bulk}$, the discrepancy for T* is not improved at all (See supplementary information). It suggests an intrinsic "Knight shift anomaly" which would be further discussed later.

On the other hand, similar electronic crossover behavior also appears in charge transport as shown in Fig. 2. Temperature-dependent resistivity shows a crossover behavior at T*. Below T*, the positive slope of resistivities for all samples shows a significant change. We try to use the cross point of two *T*-linear trends to determine T* in resistivity. It is evident that both temperature-dependent Knight shift and resistivity have a consistent crossover behavior at nearly the same T*. We would like to emphasize that such similar crossover behavior in both Knight shift and resistivity has been widely observed in *f*-electron HF materials (*4, 22*). The great similarity to *f*-electron materials strongly suggests a possible Kondo scenario for AFe$_2$As$_2$ (A = K, Rb, Cs). Indeed, such a coherent-incoherent crossover for *3d* electrons has already been proposed in theory for FeSCs (*16, 17*), and our work clearly manifests such a case. However, these results are still not enough to settle the microscopic origin of *d*-electron HF behavior in AFe$_2$As$_2$ (A = K, Rb, Cs), especially for Kondo scenario.

In order to further understand the microscopic origin of *d*-electron HF behavior, we explore the possible Kondo scaling behavior in AFe$_2$As$_2$ (A = K, Rb, Cs). As we mentioned, an intrinsic "Knight shift anomaly" is revealed by comparing the temperature-dependent Knight shift and bulk magnetic susceptibility. In Fig.3A, the K-$\chi_{bulk}$ plot for all samples shows a clear deviation from linear behavior around T*. It suggests that Knight shift is no longer scaled with bulk magnetic susceptibility below T*. Similar phenomenon has been widely found in *f*-electron HF materials (*22*). In a so-called two-fluid model of heavy fermions (*23, 24*), "Knight shift anomaly" is ascribed to the coexistence of itinerant KL and local spin liquid. The total bulk magnetic susceptibility could be expressed as $\chi = f(T)\chi_{KL} + [1-f(T)]\chi_{SL}$, where *f(T)* is an order parameter to characterize the emergent KL component. The corresponding Knight shift is $K = K_0 + A f(T)\chi_{KL} + B[1-f(T)]\chi_{SL}$, in which the spin liquid component is coupled to the probe nucleus by a transferred hyperfine coupling constant B, while the itinerant KL component is coupled by a direct hyperfine coupling A. $K_0$ is a constant offset from orbital contribution. Then, the emergent anomalous part of Knight shift is $K_a = K - K_0 - B\chi = (A-B)f(T)\chi_{KL}$, which is proportional to $\chi_{KL}$ and shows a universal T* scaling behavior in *f*-electron HF materials. In current case, we also adopt the two-fluid model to analyze the "Knight shift anomaly". In Fig.3B, the extracted $K_a$ for $AFe_2As_2$ (A = K, Rb, Cs) shows a perfect T* scaling behavior similar to the *f*-electron HF systems. In *f*-electron HF systems, $f(T) \sim (1-T/T^*)^{3/2}$ and $\chi_{KL} \sim 1+\ln(T^*/T)$. The corresponding scaling behavior of $K_a$ is proportional to $(1-T/T^*)^{3/2}(1+\ln(T^*/T))$. It shows a divergent behavior as T→0, which is regulated by the divergence in Sommerfeld coefficient ($\gamma \sim \chi_{KL}$). For $AFe_2As_2$ (A = K, Rb, Cs), the scaling curve shows a saturation behavior below a characteristic temperature $T_0$ instead of divergent behavior as T→0. Since specific heat measurement for $AFe_2As_2$ (A = K, Rb, Cs) shows a temperature-independent Sommerfeld coefficient at low temperature (See supplementary information), the saturation of $K_a$ indicates that *f(T)* reaches to upper limit and becomes to a constant below $T_0$. These results imply that the KL component in two-fluid description is completely dominated, while the local spin liquid component disappears below $T_0$. Similar saturation behavior is also observed in *f*-electron HF materials, *e.g.* $CeSn_3$ and $URu_2Si_2$ (*23, 24*). In this case, a Landau Fermi-liquid behavior would emerge at lower temperature as expected in two-fluid model (*5*). This is also consistent with the previous observation of a $T^2$ behavior in charge transport, a temperature-independent Sommerfeld coefficient and a Wilson ratio close to unity at low temperature (*20*). Therefore, these observations strongly support a consistent Kondo scenario under two-fluid model.

In addition, we also studied the low-energy magnetic fluctuation for $AFe_2As_2$ (A = K, Rb, Cs). As shown in Fig. 3C, we measured the temperature-dependent spin-lattice relaxation rate ($1/T_1$) which is related to the dynamical magnetic susceptibility $Im\chi_\perp(q,\omega_n)$ with $\frac{1}{T_1} = 2\gamma_n^2 T \sum_q A_\perp^2(q) Im\chi_\perp(q,\omega_n)/\omega_n$, where $A_\perp(q)$ is the transferred hyperfine coupling tensor perpendicular to the external field direction at $^{75}$As sites and $\omega_n = \gamma_n H$ is the NMR frequency. Usually, $1/T_1$ in Landau Fermi-liquid state follows a Korringa-type behavior with $1/T_1 T \sim N^2(E_F)$, where $N(E_F)$ is the density of state at the Fermi level and is temperature-independent. If local moments dominates the low-energy magnetic fluctuation, $1/T_1$ would obey a Curie-type behavior with $1/T_1 T \sim T^{-1}$. In Fig.3C, the temperature-dependent $1/T_1$ for $AFe_2As_2$ (A = K, Rb, Cs) exhibits a similar crossover behavior at T* as Knight shift and resistivity shown in Fig.2. Above T*, $1/T_1$ shows a temperature-independent behavior, suggesting a local-moments dominated behavior. With decreasing temperature below T*, a well-defined power-law behavior emerges with $1/T_1 \sim T^{0.75}$. Similar crossover behavior in $1/T_1$ has been observed in *f*-electron HF materials (*4*), in which the power index is different and strongly dependent on the type of quantum criticality nearby, such as 0 for $YbRh_2Si_2$ (*25*), 0.25 for $CeCoIn_5$ (*26*) and 1 for $URu_2Si_2$ (*27*). The observed $T^{0.75}$ power-law behavior in $AFe_2As_2$ (A = K, Rb, Cs) might be ascribed to a specific quantum criticality but it is still not very clear at present stage (*18, 28*), which needs further investigation. In Fig. 3D, we also plot $1/T_1$ as a function of the renormalized temperature (T/T*) for $AFe_2As_2$ (A = K, Rb, Cs), and a universal scaling behavior for $1/T_1$ is observed. It suggests that the same scaling rule in "Knight shift anomaly" also dominate the low-energy magnetic fluctuation below T*. By T* scaling, we can extract a universal $1/T_1$ formula in the coherent state, $\frac{1}{T_1} = C \times \left(\frac{T}{T^*}\right)^{0.75}$, where C = 238 ± 2 s$^{-1}$.

Recent high-resolution ARPES experiment has observed a temperature-dependent hybridization gap between d$_{xy}$ and d$_{xz}$ bands in iron chalcogenides superconductors (*29*). Meanwhile, a considerable band renormalization effect was also revealed for d$_{xy}$ band. This result might be treated as a possible microscopic picture for the Kondo scenario proposed in present study. The d$_{xy}$ band would play the role of local moments as *f*-electrons did in HF materials. The first principle calculation also supports the local character for d$_{xy}$ band in KFe$_2$As$_2$ (*20*), which is consistent with quantum oscillation (*14*) and APRES experiment (*15*). Further high-resolution ARPES experiment to explore the hybridization gap and Kondo coherent peak would be required to complete the microscopic picture for Kondo scenario in AFe$_2$As$_2$ (A = K, Rb, Cs).

In various *f*-electron HF superconductors, the T$_c$ is always less than the order of 0.1T* (*30*). In Fig. 4, we show a summary plot for the relationship among T$_c$, T* and γ. Although T$_c$ is hard to exceed the order of 0.1T*, it's very clear that higher T* favors higher T$_c$. In Kondo lattice systems, T* follows an empirical scaling behavior with T* = cJ$^2$ρ (*4*), in which J is the local Kondo coupling, and ρ is the density of states of the conduction electrons coupled to the local spins and c is a constant determined by the details of the hybridization and the conduction electron Fermi surface. The relationship between T$_c$ and T* (in Fig. 4) suggests that the unconventional superconductivity in HF superconductors is correlated to local magnetic coupling (*30*). Based on the present NMR results, it's reasonable to classify the AFe$_2$As$_2$ family (A = K, Rb, Cs) as new HF superconductors without *f*-electrons. As expected under Kondo scenario, the AFe$_2$As$_2$ family (A = K, Rb, Cs) exhibit a consistent tendency with the *f*-electron HF superconductors as shown in Fig.4. The systematic evolution of local magnetic coupling due to modification of Fe-Fe distance by the alkali metal with different ionic radius might be responsible for the variation of T* in AFe$_2$As$_2$ family (A = K, Rb, Cs) (*28*). On the other hand, T* is also expected to be related to Sommerfeld coefficient with γ$^{-1}$ ~ T* in two-fluid model of heavy fermions as KL component becomes dominated at low temperature (*5*). As shown in Fig.3B, the observed KL component in AFe$_2$As$_2$ family (A = K, Rb, Cs) also becomes dominated below T$_0$. Therefore, we would expect that the Sommerfeld coefficients at low temperature should be correlated to T*. As shown in Fig. 4, the Sommerfeld coefficients of AFe$_2$As$_2$ family (A = K, Rb, Cs) do follow the same T* dependent behavior as the *f*-electron HF superconductors. All these results further support AFe$_2$As$_2$ family (A = K, Rb, Cs) as the first material realization of *d*-electron HF superconductors under the frame of Kondo scenario.

A novel route to achieve high-T$_c$ superconductivity in HF superconductors is to search for high-T* HF materials (*30*). Here, in *d*-electron HF materials, the T* would be expected higher than that of *f*-electron HF materials due to stronger hybridization effect. Whether the newly discovered high-T$_c$ superconductivity in FeSCs could be understood in this way would be an interesting issue. This would be helpful to build a universal picture for unconventional superconductivity in both FeSCs and HF superconductors.

**Methods:**

**1. Sample growth**

High-quality $AFe_2As_2$ (A = K, Rb, Cs) single crystals are grown by the self-flux technique. The K/Rb/Cs chunks, Fe and As powders were weighted according to the ratio of K/Rb/Cs:Fe:As=6:1:6. Typically, 1.5 grams of the mixture of Fe and As powders were loaded into a 10-mm diameter alumina crucible, and freshly cut K/Rb/Cs pieces were placed on top of the mixture. Then the alumina crucible with a lid was sealed in a stainless steel container assembly. The whole preparation process was carried out in a glove box in which high pure argon atmosphere was filled ($O_2$ content is less than 1 ppm). The sealed stainless steel assembly was then sealed inside an evacuated quartz tube. The quartz tube was placed in a box furnace and slowly heated up to 200 °C. It was kept at 200 °C for 400 minutes, which allows full reaction of K/Rb/Cs and the mixture. Then the sample was heated up to 950 °C in 10 hours. The temperature was kept still for 10 hours and then slowly cooled to 550 °C at a rate of 3 °C/h. After cooling down to room temperature by switching off the furnace, shiny platelike crystals can be easily picked up from the alumina crucible. The single crystals are stable in air or alcohol for several days.

**2. Nuclear magnetic resonance**

Standard NMR spin-echo techniques were used with a commercial NMR spectrometer from Thamway Co. Ltd. The external magnetic field was generated by a 12 Tesla magnet from Oxford Instruments. The NMR coils were made by copper wires. The $^{27}Al$ NMR signal from 0.8 μm-thick aluminum foils (99.1% purity) was used to calibrate the external field. $^{75}As$ NMR spectra were obtained by sweeping the frequency at fixed magnetic field values and then integrating the spin-echo signal for all frequency values. Spin-lattice relaxation time ($T_1$) measurement was extracted from fitting spin-echo decay with formula $I(t) = I_0 + I_1 \times \left[0.1 \times \exp(-(\frac{t}{T_1})^r) + 0.9 \times \exp(-(\frac{6t}{T_1})^r)\right]$ for NMR measurement on central line and $I(t) = I_0 + I_1 \times \exp(-(\frac{3t}{T_1})^r)$ for NQR measurement.

**3. Electronic transport, magnetization and XRD characterization**

The magnetic susceptibility was measured using a vibrating sample magnetometer (VSM) (Quantum Design). The direct current (dc) resistivity was measured by the conventional four-probe method using the PPMS-9T (Quantum Design). XRD was performed on a SmartLab-9 diffractometer (Rikagu) from 10° to 70° with a scanning rate of 4° per minute.

**References:**

1. Monthoux, P., Pines, D. & Lonzarich, G. G. Superconductivity without phonons. *Nature* **450**, 1177 (2007).
2. Gegenwart, P., Si, Q. & Steglich, F. Quantum criticality in heavy-fermion metals. *Nat. Phys.* **4**, 186-197 (2008).
3. Nakatsuji, S., Pines, D. & Fisk, Z. Two Fluid Description of the Kondo Lattice. *Phys. Rev. Lett.* **92**, 016401 (2004).
4. Yang, Y.-F. *et al.* Scaling the Kondo lattice. *Nature* **454**, 611 (2008).
5. Yang, Y.-F. & Pines, D. Emergent states in heavy-electron materials. *Proc. Natl. Acad. Sci. USA* **45**, E3060–E3066 (2012).
6. Anisimov, V. I. *et al.* Electronic Structure of the Heavy Fermion Metal $LiV_2O_4$. *Phys. Rev. Lett.* **83**, 364–367 (1999).
7. Arita, R., Held, K., Lukoyanov, A. V. & Anisomov, V. I. Doped Mott Insulator as the Origin of Heavy-Fermion Behavior in $LiV_2O_4$. *Phys. Rev. Lett.* **98**, 166402 (2007).
8. Hopkinson, J. & Coleman, P. $LiV_2O_4$: Frustration Induced Heavy Fermion Metal. *Phys. Rev. Lett.* **89**, 267201 (2002).
9. Yamashita, Y. & Ueda, K. Spin-orbital fluctuations and a large mass enhancement in $LiV_2O_4$. *Phys. Rev. B* **67**, 195107 (2003).
10. Kim, J. S. *et al.* Specific heat in $KFe_2As_2$ in zero and applied magnetic field. *Phys. Rev. B* **83**, 172502 (2011).
11. Zhang, Z. *et al.* Heat transport in $RbFe_2As_2$ single crystals: Evidence for nodal superconducting gap. *Phys. Rev. B* **91**, 024502 (2015).
12. Wang, A. F. *et al.* Calorimetric study of single-crystal $CsFe_2As_2$. *Phys. Rev. B* **87**, 214509 (2013).
13. Hardy, H. *et al.* Doping evolution of superconducting gaps and electronic densities of states in $Ba(Fe_{1-x}Co_x)_2As_2$ iron pnictides. Europhys. *Lett.* **91**, 47008 (2010).
14. Terashima, T. *et al.* Fermi surface in $KFe_2As_2$ determined via de Haas–van Alphen oscillation measurements. *Phys. Rev. B* **87**, 224512 (2013).
15. T. Sato. *et al.* Band Structure and Fermi Surface of an Extremely Overdoped Iron-Based Superconductor $KFe_2As_2$. *Phys. Rev. Lett.* **103**, 047002 (2009).
16. Haule, K. & Kotliar, G. Coherence–incoherence crossover in the normal state of iron oxypnictides and importance of Hund's rule coupling. *New J. Phys.* **11**, 025021 (2009).
17. Yin, Z. P., Haule, K. & Kotliar, G. Kinetic frustration and the nature of the magnetic and paramagnetic states in iron pnictides and iron chalcogenides. *Nat. Mater.* **10**, 932 (2011).
18. Eilers, F. *et al.* Quantum criticality in $AFe_2As_2$ with A = K, Rb, and Cs suppresses superconductivity. *arXiv*:1510.01857 (2015).
19. Dai, P. C., Hu, J. P. & Dagotto, E. Magnetism and its microscopic origin in iron-based high-temperature superconductors. *Nat. Phys.* **8**, 709 (2012).
20. Hardy, F. *et al.* Evidence of Strong Correlations and Coherence-Incoherence Crossover in the Iron Pnictide Superconductor $KFe_2As_2$. *Phys. Rev. Lett.* **111**, 027002 (2013).
21. Fukazawa, H. *et al.* Possible Multiple Gap Superconductivity with Line Nodes in Heavily Hole-Doped Superconductor $KFe_2As_2$ Studied by $^{75}As$ Nuclear Quadrupole Resonance and Specific Heat. *J. Phys. Soc. Jpn.* **78**, 083712 (2009).
22. Curro, N. J. Nuclear magnetic resonance in the heavy fermion superconductors. *Rep. Prog. Phys.* **72**, 026502 (2009).

23. Curro, N. J. *et al.* Scaling in the emergent behavior of heavy-electron materials. *Phys. Rev. B* **70**, 235117 (2004).
24. Yang, Y. -F. & Pines, D. Universal Behavior in Heavy-Electron Materials. *Phys. Rev. Lett.* **100**, 096404 (2008).
25. Ishida, K. *et al.* YbRh$_2$Si$_2$: Spin Fluctuations in the Vicinity of a Quantum Critical Point at Low Magnetic Field. *Phys. Rev. Lett.* **89**, 107202 (2002).
26. Kohori, Y. *et al.* NMR and NQR studies of the heavy fermion superconductors CeTIn$_5$ (T=Co and Ir). *Phys. Rev. B* **64**, 134526 (2001).
27. Kohara, T. *et al.* Numerical renormalization study of a tunneling atom with Kondo spin. *Solid State Communication.* **59**, 603-606 (1986).
28. Mizukami, Y. *et al.* Evolution of quasiparticle excitations with critical mass enhancement in superconducting AFe$_2$As$_2$ (A = K, Rb, and Cs). *arXiv*:1510.02273 (2015).
29. Yi, M. *et al.* Observation of universal strong orbital-dependent correlation effects in iron chalcogenides. *Nat. Comm.* **6**, 7777 (2014).
30. Pines, D. Finding new superconductors: the spin-fluctuation gateway to high $T_c$ and possible room temperature superconductivity. *J. Phys. Chem. B* **117**, 13145(2013).

**Acknowledgments**

The authors are grateful for the stimulating discussions with Z. P. Yin, G. Kotliar, Y.-F. Yang, G. M. Zhang, D. H. Lee, F. C. Zhang, M.-H. Julien, S. Y. Li and Z. Sun. This work is supported by the National Natural Science Foundation of China, the "Strategic Priority Research Program (B)" of the Chinese Academy of Sciences, the Fundamental Research Funds for the Central Universities and the Chinese Academy of Sciences. T. W. acknowledges the Recruitment Program of Global Experts and the CAS Hundred Talent Program.

**Author contributions**

T.W. and X.H.C. conceived and designed the experiments. T.W. and X.H.C. are responsible for the infrastructure and project direction. Y.P.W. and D.Z. performed NMR measurements with the assistance from T.W.. A.F.W., N.Z.W., Z.J.X., X.G.L. and X.H.C. performed sample growth, bulk magnetic susceptibility and transport characterization. Y.P.W. and T.W. analyzed and interpreted the NMR results. T.W. and X.H.C. wrote the manuscript. All authors discussed the results and commented on the manuscript.

**Additional Information**

The authors declare no competing financial interests. Correspondence and requests for materials should be addressed to T. W. (wutao@ustc.edu.cn) or X.H.C. (chenxh@ustc.edu.cn).

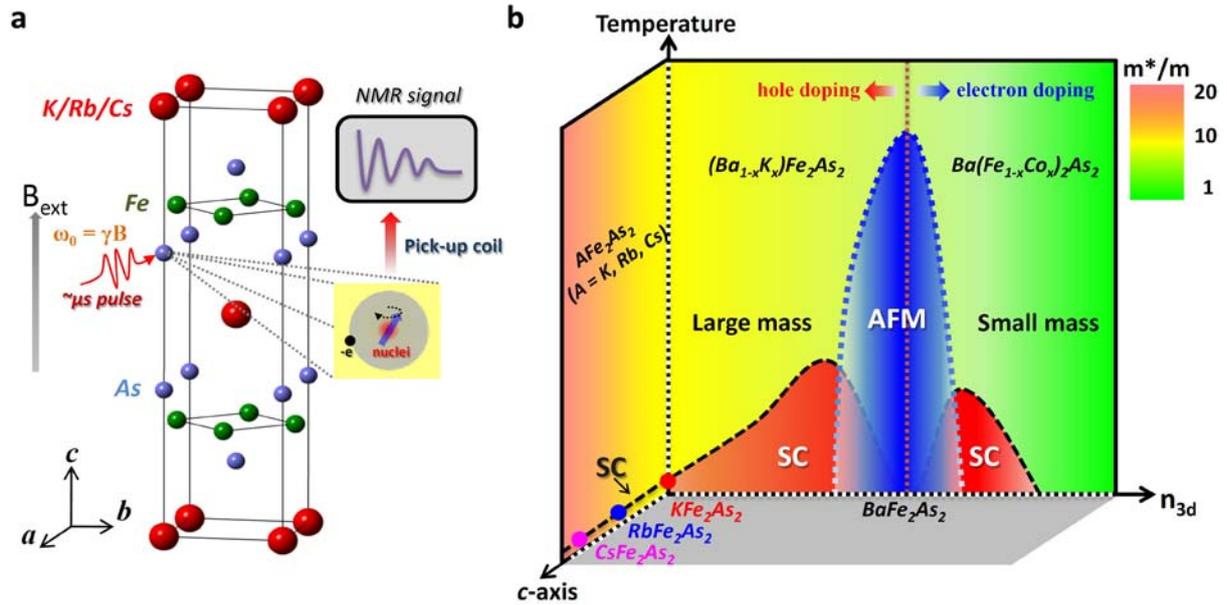

**Fig.1. Crystal structure, sketch of NMR experiment and schematic phase diagram of FeSCs with 122 structure. a**. Crystal structure of heavily hole-doped FeSCs AFe$_2$As$_2$ (A = K, Rb, Cs) and sketch of NMR experiment. The local symmetry of the As site is tetragonal with four nearest-neighbour Fe atoms. We performed NMR measurements on $^{75}$As nuclei. The nuclei spin number of $^{75}$As is 3/2. The applied external magnetic field (B$_{ext}$) in this study is along c axis of crystal lattice. ω$_0$ is the Larmor frequency of nuclei spin under B$_{ext}$. γ is gyromagnetic ratio for nuclei spin. After a pulse with several microseconds, a NMR signal is observed through pick-up coil. **b**. Schematic phase diagram of FeSCs with 122 structure tuned by the number of 3d electrons (n$_{3d}$) at Fe sites. The color in phase diagram depicts effective mass m* relative to that from band theory. SC denotes superconducting phase. AFM denotes antiferromagnetic phase.

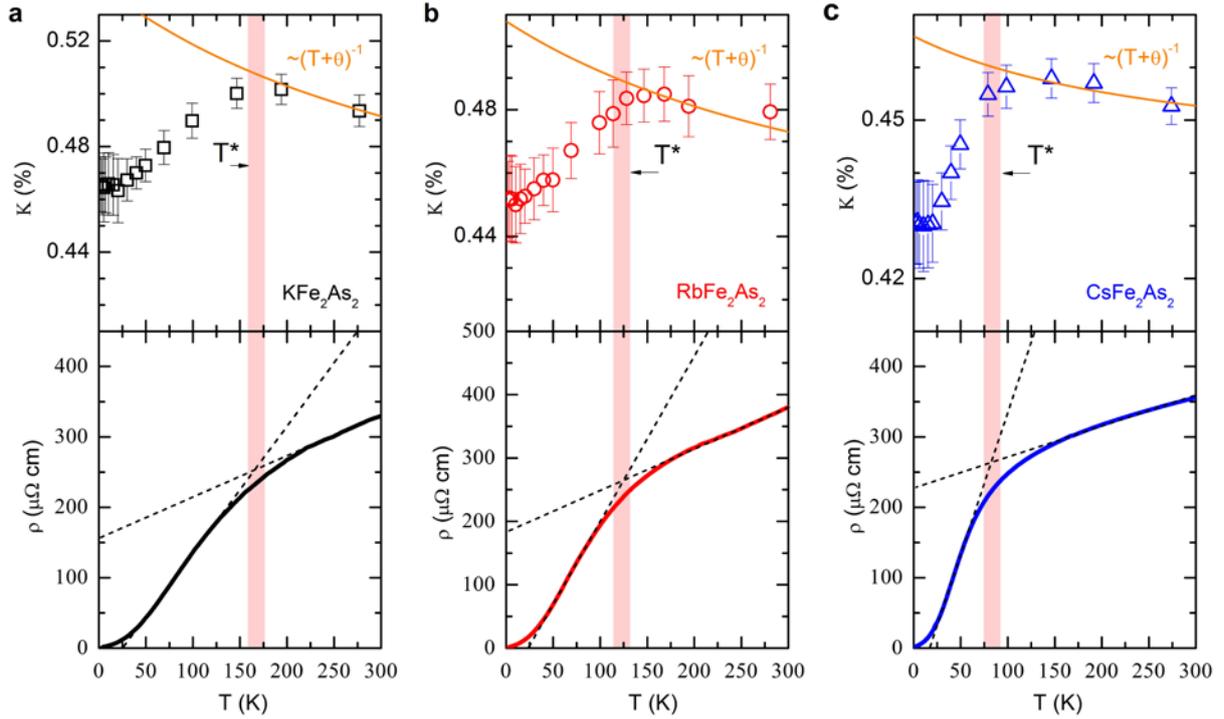

**Fig. 2. Universal coherent-incoherent crossover behavior in Knight shift and resistivity of $AFe_2As_2$ (A = K, Rb, Cs).** Top panels: Temperature-dependent Knight shift (K) for **a**. $KFe_2As_2$; **b**. $RbFe_2As_2$; **c**. $CsFe_2As_2$. External magnetic field ($H_{ext}$) was applied parallel to the c-axis with a magnitude of 12 Tesla. Error bars were determined by the full width at half maximum (FWHM) of central line. Solid lines represent Curie-Weiss fittings for K(T) above T*. The values of θ are extracted from a similar Curie-Weiss fitting on bulk magnetic susceptibility (See supplementary information). Bottom panels: Temperature- dependent resistivities for **a**. $KFe_2As_2$; **b**. $RbFe_2As_2$; **c**. $CsFe_2As_2$. The dash lines are two tentative T-linear trends to extract the value of T* from resistivity. The bold pink lines mark nearly the same T* determined by Knight shift and resistivity.

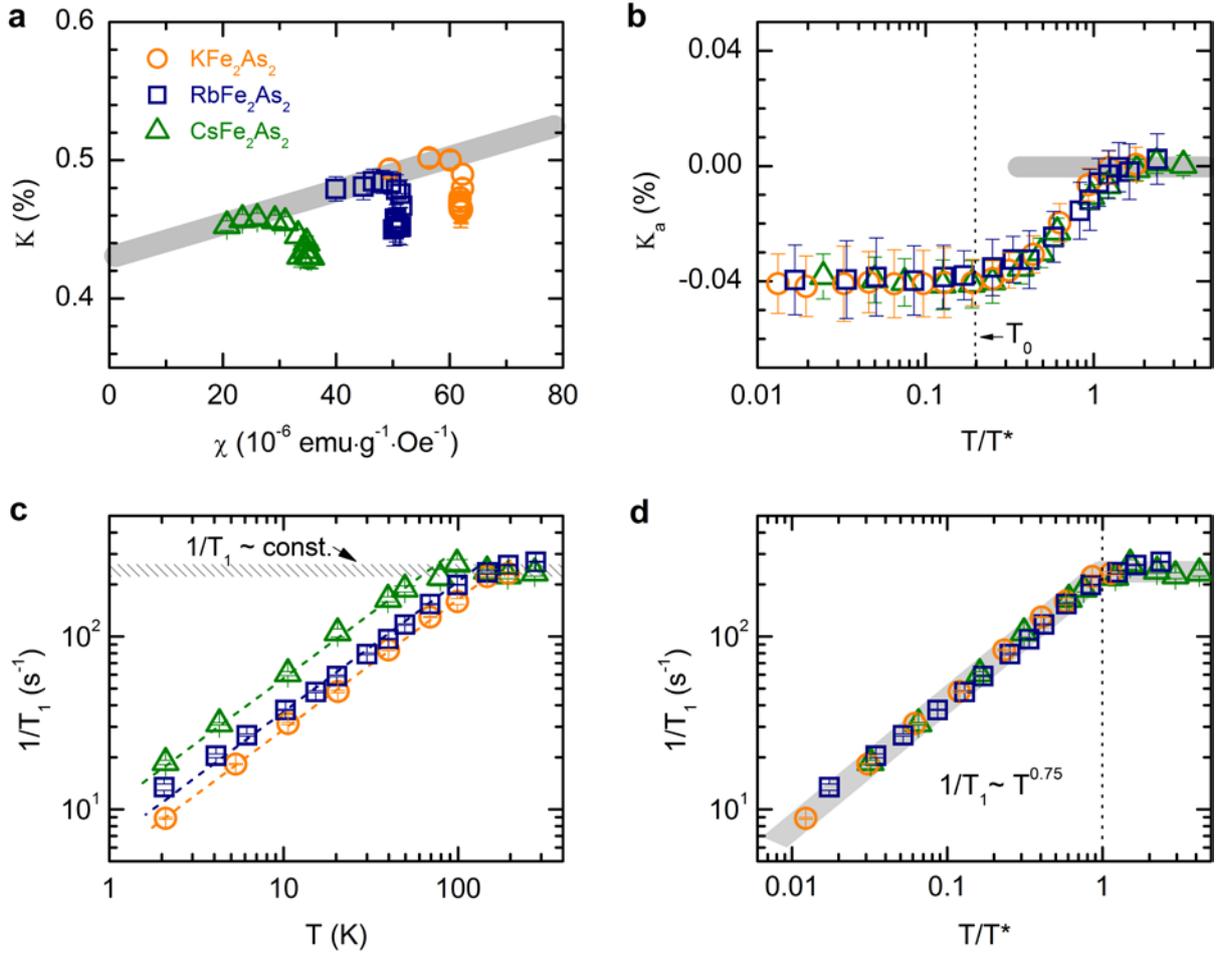

**Fig. 3. Emergent Kondo Scaling in $AFe_2As_2$ (A = K, Rb, Cs). a.** $K$-$\chi$ plot for $AFe_2As_2$ (A = K, Rb, Cs). The Knight shift data used here is the same as that in Fig. 2. The grey bold line is guide for eyes. It points out a clear deviation from linear behavior around T*. Here, Curie-Weiss tail contribution on bulk magnetic susceptibility ($\chi$) has already been subtracted (See supplementary information). **b.** Scaling behavior for anomalous Knight shift ($K_a$). The definition of $K_a$ is described in the main text. The dotted line represents another characteristic temperature $T_0$, below which the anomalous Knight shift becomes saturated. Here, $T_0$ is about 0.2T*. **c.** Temperature-dependent $1/T_1$ for $AFe_2As_2$ family (A = K, Rb, Cs). The experimental details for $1/T_1$ see supplementary information. Above T*, $1/T_1$ exhibits a temperature independent behavior. Below T*, all samples exhibit a similar temperature dependence with $1/T_1 \sim T^{0.75}$. Error bars were determined by least squares fits to the nuclear relaxation curves. **d.** Scaling behavior for temperature-dependent $1/T_1$ in $AFe_2As_2$ (A = K, Rb, Cs). The grey bold line represents a universal scaling behavior with $1/T_1 \sim (T/T^*)^{0.75}$.

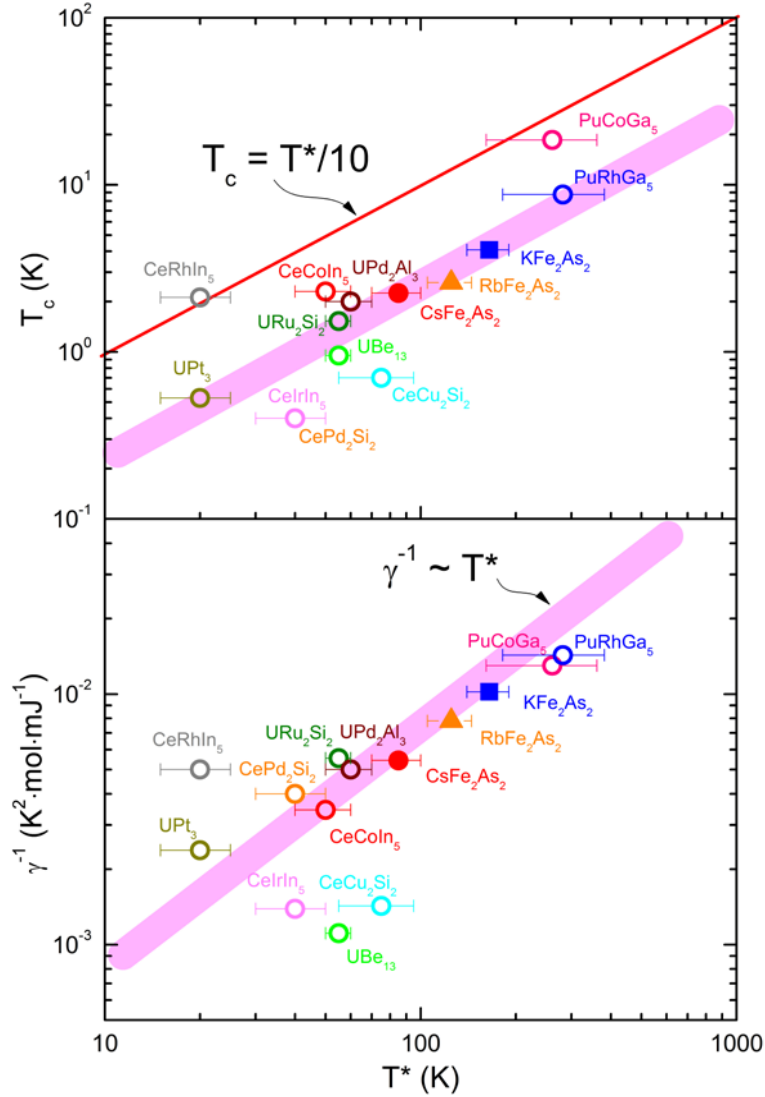

**Fig. 4. Universal relationship of $T_c$ and Sommerfeld coefficient with T* in heavy fermion superconductors.** Top panel: the correlation between the superconducting temperature $T_c$ and the characteristic Kondo temperature T* for *f*-electron HF superconductors and $AFe_2As_2$ (A = K, Rb, Cs). The red line is an empirical upper limit for the maximum of $T_c$ for f-electron HF superconductors with a given T*. The grey bold line is a guide for eyes. It represents a universal tendency for the evolution of $T_c$ with T*. The $T_c$ values for $AFe_2As_2$ (A = K, Rb, Cs) are 4.08 K, 2.60 K, and 2.25 K, respectively (See supplementary information). Bottom panel: the correlation between Sommerfeld coefficient ($\gamma$) and T* for *f*-electron HF superconductors and $AFe_2As_2$ (A = K, Rb, Cs). The grey bold line represents a universal relationship with $\gamma^{-1} \sim$ T*. It's clear that the reciprocal of Sommerfeld coefficient in $AFe_2As_2$ (A = K, Rb, Cs) follows the grey bold line for most of *f*-electron HF superconductors. All data used in Fig. 4 is listed in supplementary information.

# Supplementary Information:

# Emergent Kondo scaling in iron-based superconductors $AFe_2As_2$ (A = K, Rb, Cs)

Y. P. Wu[1], D. Zhao[1], A. F. Wang[1], N. Z. Wang[1], Z. J. Xiang[1], X. G. Luo[1,2,4], T. Wu[1,2,4] and X. H. Chen[1,2,3,4]

1. Hefei National Laboratory for Physical Sciences at Microscale and Department of Physics, University of Science and Technology of China, Hefei, Anhui 230026, China

2. Key Laboratory of Strongly-coupled Quantum Matter Physics, Chinese Academy of Sciences, School of Physical Sciences, University of Science and Technology of China, Hefei, Anhui 230026, China

3. High Magnetic Field Laboratory, Chinese Academy of Sciences, Hefei, Anhui 230031, China

4. Collaborative Innovation Center of Advanced Microstructures, Nanjing University, Nanjing 210093, China

**Outline:**

1. XRD characterization for $AFe_2As_2$ (A = K, Rb, Cs)

2. Superconducting transition in temperature-dependent resistivity for $AFe_2As_2$ (A = K, Rb, Cs)

3. $^{75}$As NMR spectrum for $AFe_2As_2$ (A = K, Rb, Cs)

4. Temperature-dependent bulk magnetic susceptibility for $AFe_2As_2$ (A = K, Rb Cs)

5. $T_1$ decay for NMR and NQR measurement in $CsFe_2As_2$

6. Temperature-dependent stretch in $T_1$ fitting procedure for $KFe_2As_2$ and $RbFe_2As_2$

7. List of T*, Tc and γ for different heavy fermion superconductors

8. Specific heat measurement on $CsFe_2As_2$

1. **XRD characterization for $AFe_2As_2$ (A = K, Rb, Cs)**

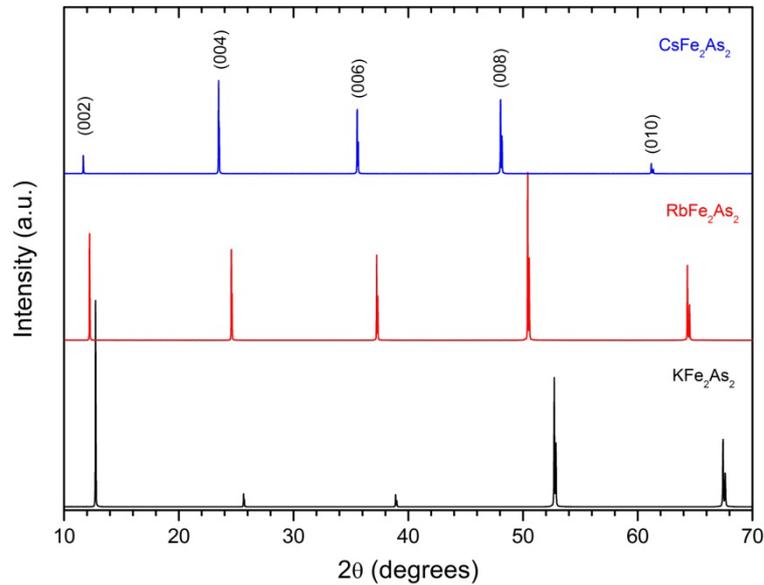

**Fig. S1. XRD patterns for $AFe_2As_2$ (A = K, Rb, Cs).** The c-axis lattice parameter is estimated to be c = 13.88 Å for $KFe_2As_2$, c = 14.45 Å for $RbFe_2As_2$, c = 15.13 Å for $CsFe_2As_2$, respectively. Only (00l) reflections can be recognized, indicating that the crystals are well orientated along the c axis.

2. **Superconducting transition in temperature-dependent resistivity for $AFe_2As_2$ (A = K, Rb, Cs)**

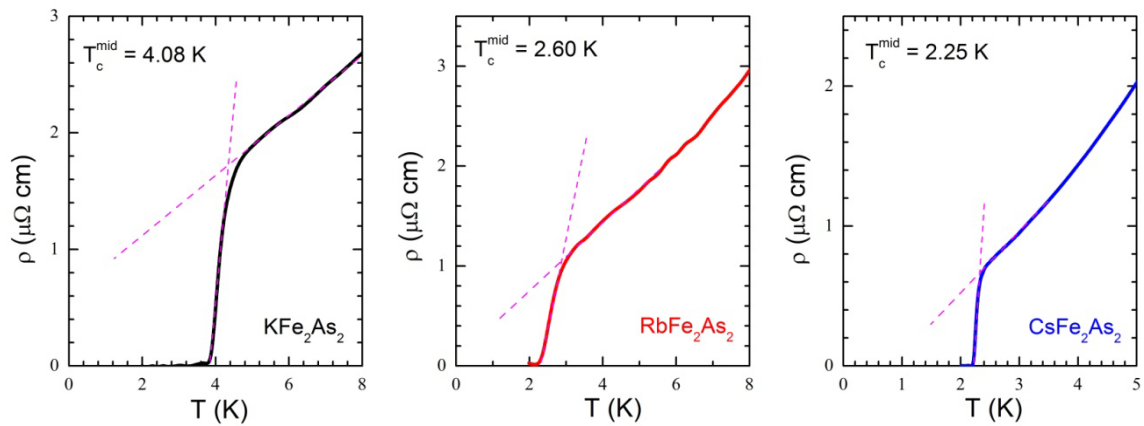

**Fig. S2. Superconducting transition in temperature-dependent resistivity for $AFe_2As_2$ (A = K, Rb, Cs).** The superconducting transition temperature $T_c$ is defined by the mid-point of the superconducting transition. In this way, the $T_c$ and transition width for $AFe_2As_2$ (A = K, Rb, Cs) is 4.08 K and 0.60 K for $KFe_2As_2$, 2.60 K and 0.70 K for $RbFe_2As_2$, 2.25 K and 0.10 K for $CsFe_2As_2$, respectively. The extremely narrow superconducting transition width for $CsFe_2As_2$ is quite consistent with NMR satellite shown in Fig. S3, in which $CsFe_2As_2$ shows a much narrower satellite than $KFe_2As_2$ and $RbFe_2As_2$. This suggests a better sample quality

for CsFe$_2$As$_2$.

## 3. $^{75}$As NMR spectrum for AFe$_2$As$_2$ (A = K, Rb, Cs)

The central line and satellite of NMR spectrum for AFe$_2$As$_2$ (A = K, Rb, Cs) are shown in Fig. S3. There is no significant change for NMR spectrum for AFe$_2$As$_2$ (A = K, Rb, Cs).

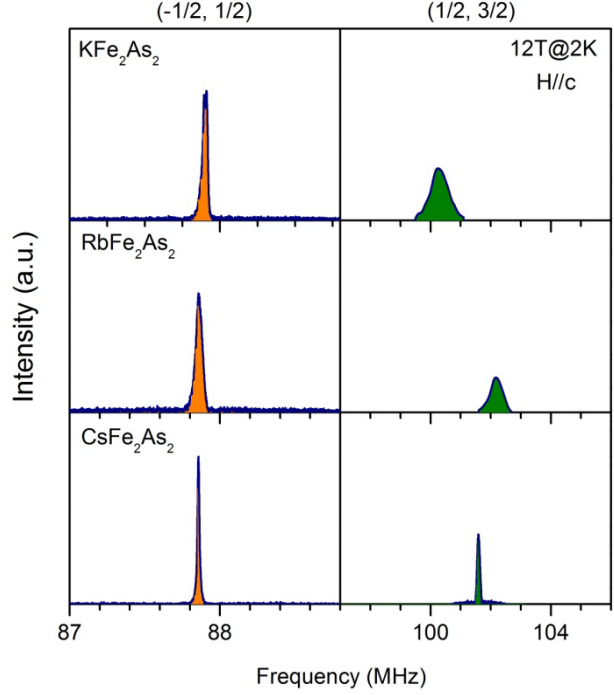

**Fig. S3. $^{75}$As NMR spectrum for AFe$_2$As$_2$ (A = K, Rb, Cs).** The nuclear spin (I) for $^{75}$As is 3/2. (-1/2, 1/2) represents the central transition of $^{75}$As NMR. (-3/2, -1/2) and (1/2, 3/2) represent satellite transitions of $^{75}$As NMR. Due to quadrupole interaction, the central and satellite transitions show different frequency from central transition with splitting frequency $\upsilon_{\alpha\alpha} = \frac{3e^2 QV_{\alpha\alpha}}{2I(I-1)h}$ ($\alpha$ = a, b, c), Q is electric quadrupole moment for $^{75}$As nuclei and V$_{\alpha\alpha}$ is electric field gradient on $^{75}$As sites. The measuring temperature is at 2 K and the applied external field is 12 T along c axis.

## 4. Temperature-dependent bulk magnetic susceptibility for AFe$_2$As$_2$ (A = K, Rb Cs)

As shown in Fig. S4, the temperature-dependent bulk magnetic susceptibilities show a Curie-Weiss type behavior ($\chi = \frac{C}{T+\theta}$) above T*, consistent with a local-moment dominated behavior as observed in Knight shift measurement. An additional Curie-Weiss tail at the lowest temperature is also observed in bulk magnetic susceptibility, especially for KFe$_2$As$_2$ and RbFe$_2$As$_2$. This is an clear evidence for impurity effect. We try to subtract the Curie-Weiss tail from the bulk magnetic susceptibility. For CsFe$_2$As$_2$, the Curie-Weiss tail is not very clear and we adopt a M-H method to subtract impurity effect as shown in Fig. S5.

The impurity contribution extracted from this method is also proved to be Curie-Weiss type, which is consistent with that in KFe$_2$As$_2$ and RbFe$_2$As$_2$.

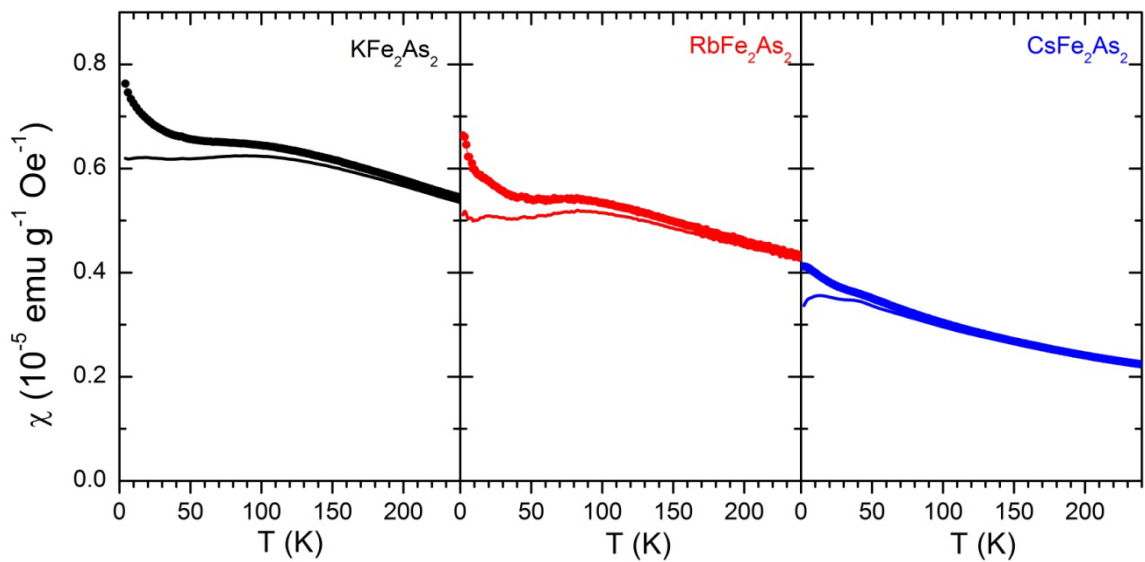

**Fig. S4. Temperature-dependent bulk magnetic susceptibility for AFe$_2$As$_2$ family (A = K, Rb, Cs).** The measurement is conducted with external field (H$_{ext}$) along c-axis. The solid line is the extracted intrinsic bulk magnetic susceptibility with subtracting a Curie-Weiss tail contribution.

**Table S1. Curie-Weiss fitting parameters for AFe$_2$As$_2$ (A = K, Rb, Cs).** The Curie-Weiss fitting procedure is conducted above 150 K for all samples.

|  | θ (K) | μ$_{eff}$ (μ$_B$ per Fe site)* |
|---|---|---|
| KFe$_2$As$_2$ | 427 | 1.48 |
| RbFe$_2$As$_2$ | 455 | 1.45 |
| CsFe$_2$As$_2$ | 290 | 0.97 |

*The effective magnetic moment is calculated from $C$ in Curie-Weiss formula.

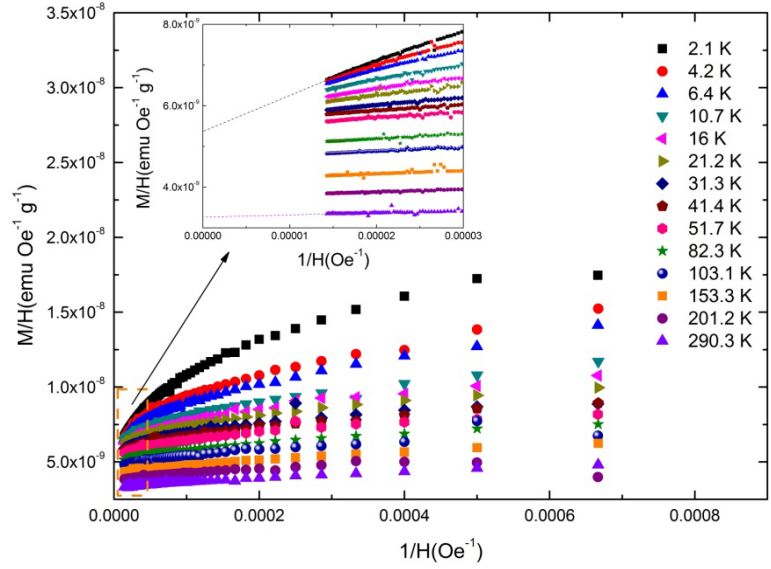

**Fig. S5. Field-dependent magnetization for CsFe$_2$As$_2$.** The measurement is conducted with external field (H$_{ext}$) along c-axis. A linear extrapolation was used to extract intrinsic bulk magnetic susceptibility as shown in the inset plot.

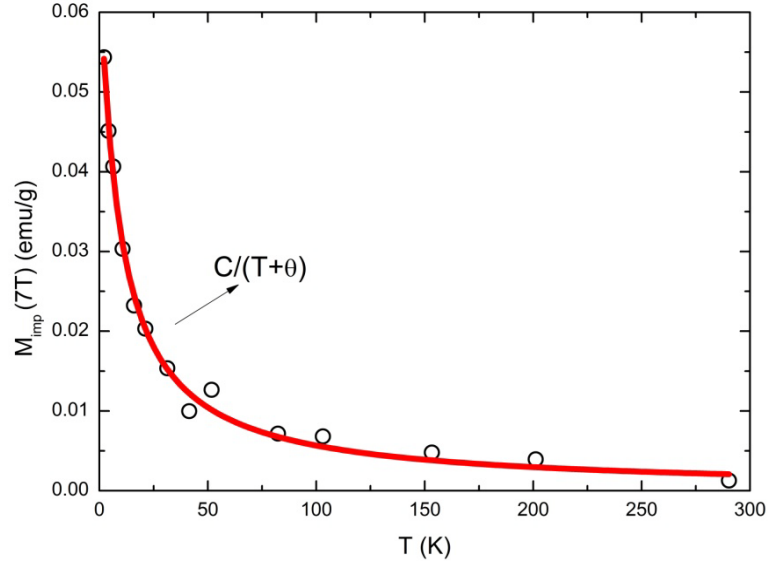

**Fig. S6. Curie-Weiss tail in CsFe$_2$As$_2$.** The impurity contribution on bulk magnetic susceptibility was extracted from M-T curve in Fig. S4 by subtracting intrinsic bulk magnetic susceptibility extracted from Fig. S5. It's evident that the impurity contribution also follows a Curie-Weiss behavior.

## 5. T$_1$ decay for NMR and NQR measurement in CsFe$_2$As$_2$

Below 20 K, there is a stretched T$_1$ decay observed for CsFe$_2$As$_2$ samples. Such stretched T$_1$

decay could be fitted by two-component $T_1$ formula instead of single component formula as shown in Fig. S7. We found that such two-component behavior in $T_1$ decay disappears in $T_1$ decay measured by NQR. It suggests a possible field-induced stretch in $T_1$. Usually, this could be ascribed to impurity effect. However, considering the narrow linewidth of $CsFe_2As_2$ as shown in Fig. S8, we believe that above field-induced two-component behavior in $T_1$ decay should be an intrinsic property. Here, we would leave such issue in future study and only take the value of $T_1$ from NQR measurement at low temperature in Fig. 3. Since the principle axis of the electric field gradient tensors in tetragonal structure is along the c-axis, it is expected that the $T_1$ from NMR measurement for external magnetic field along c-axis are equivalent to that from NQR measurement (*1*). In fact, the NMR and NQR measurements do give a same $T_1$ in present study as the two-component behavior disappears above 20 K.

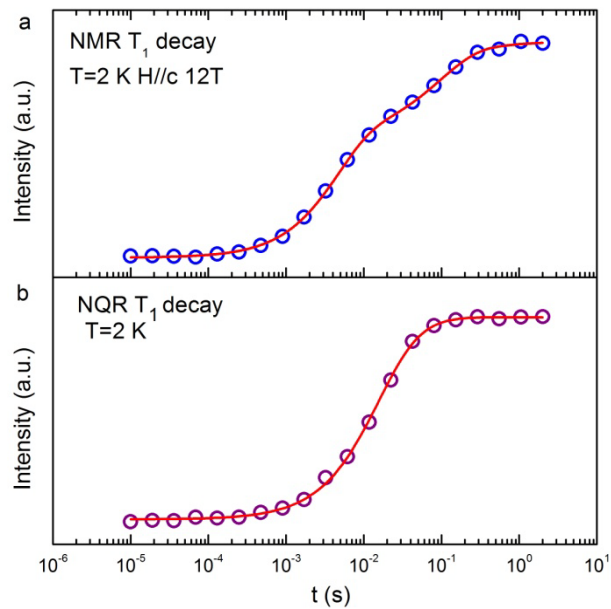

**Fig. S7. $T_1$ decay of NMR and NQR measurement for $CsFe_2As_2$.** **a**) The two-component fittings for $T_1$ decay for NMR measurement under magnetic field of 12 T at 2 K. The open circles stand for experimental data. The red solid lines stand two-component fitting result with formula: $I(t) = I_0 + I_s \times \left[0.1 \times \exp(-(\frac{t}{T_{1s}})) + 0.9 \times \exp(-(\frac{6t}{T_{1s}}))\right] + I_L \times \left[0.1 \times \exp(-(\frac{t}{T_{1L}})) + 0.9 \times \exp(-(\frac{6t}{T_{1L}}))\right]$. **b**) A perfect single-component fitting for $T_1$ decay for NQR measurement under zero field at 2 K. The open circles stand for experimental data. The red solid line is single-component fitting result with formula: $I(t) = I_0 + I_1 \times \exp(-(\frac{3t}{T_1})^r)$.

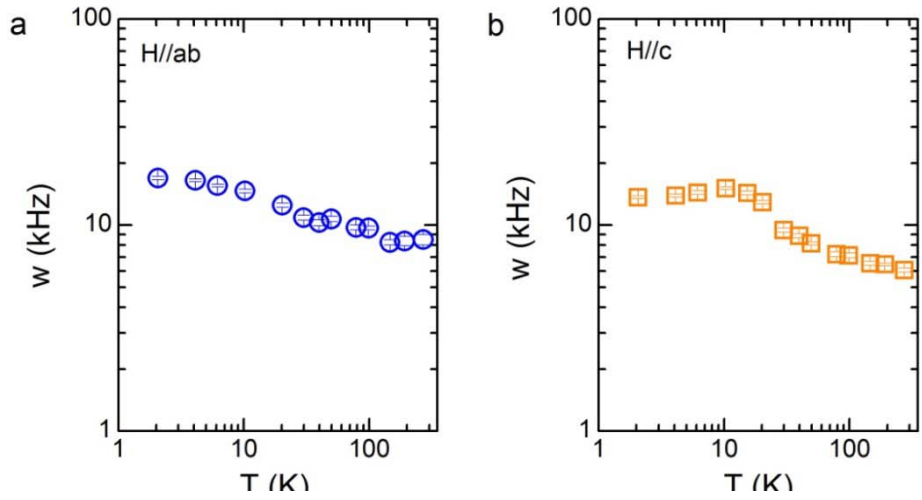

**Fig. S8. Temperature-dependent linewidth of central line in $^{75}$As NMR spectrum of CsFe$_2$As$_2$ single crystal. a**) Temperature-dependent linewidth of central transition with H//ab = 12 T. **b**) Temperature-dependent linewidth of central transition with H//c = 12 T.

## 6. Temperature-dependent stretch in $T_1$ fitting procedure for KFe$_2$As$_2$ and RbFe$_2$As$_2$

Although we observed a stretched $T_1$ decay for KFe$_2$As$_2$ and RbFe$_2$As$_2$, there is still no clear two-component feature as that in CsFe$_2$As$_2$. We only use stretched single $T_1$ formula to fit the $T_1$ decay for KFe$_2$As$_2$ and RbFe$_2$As$_2$. As shown in Fig. S9, a clear deviation from unity in stretch appears below 20 K with a smaller value of the stretch. Usually, such smaller stretch could be ascribed to inhomogeneity or a distribution of $T_1$ values (*2*). Above 20 K, the stretch recovers to a constant value close to unity.

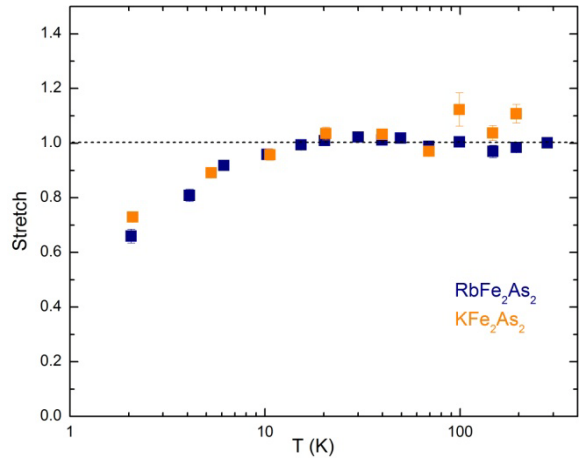

**Fig. S9. Temperature-dependent stretch in $T_1$ fitting for KFe$_2$As$_2$ and RbFe$_2$As$_2$.** The NMR measurement is conducted with H//c = 12T. The single component formula is $I(t) = I_0 + I_1 \times \left[0.1 \times \exp(-(\frac{t}{T_1})^r) + 0.9 \times \exp(-(\frac{6t}{T_1})^r)\right]$, where r represents stretch.

## 7. List of T*, Tc and γ for different heavy fermion superconductors

|          | T* (K)      | Tc (K)    | γ (mJ/mol K$^2$) |
|----------|-------------|-----------|------------------|
| CeRhIn$_5$   | 20±5 [3]    | 2.12 [5]  | 200 [7]   |
| CeCoIn$_5$   | 50±10 [3]   | 2.3 [5]   | 290 [7]   |
| CePd$_2$Si$_2$ | 40±10 [3]   | 0.4 [5]   | 250 [8]   |
| CeIrIn$_5$   | 40±10 [3]   | 0.4 [5]   | 720 [5]   |
| CeCu$_2$Si$_2$ | 75±20 [3]   | 0.7 [5]   | 700 [7]   |
| URu$_2$Si$_2$ | 55±5 [3]    | 1.53 [5]  | 180 [12]  |
| UPd$_2$Al$_3$ | 60±10 [3]   | 2 [5]     | 200 [5]   |
| UBe$_{13}$    | 55±5 [3]    | 0.95 [5]  | 900 [7]   |
| UPt$_3$       | 20±5 [3]    | 0.53 [5]  | 420 [7]   |
| PuCoGa$_5$    | 260±100 [4] | 18.5 [4]  | 77 [5]    |
| PuRhGa$_5$    | 280±100 [4] | 8.7 [4]   | 70 [5]    |
| KFe$_2$As$_2$ | 165±25 *    | 4.08 *    | 98 [9]    |
| RbFe$_2$As$_2$ | 125±20 *    | 2.60 *    | 127 [10]  |
| CsFe$_2$As$_2$ | 85±15 *     | 2.25 *    | 184 [11]  |

*from present work

## 8. Specific heat measurement on CsFe$_2$As$_2$

As shown in Fig. S9, specific heat measurement on CsFe$_2$As$_2$ sample suggests a temperature-independent Sommerfeld coefficient at low temperature. There is no clear divergent behavior as that in heavy fermion materials. This is consistent with our previous specific heat measurement (*11*). Specific heat result definitely supports a Landau Fermi liquid at low temperature. Here, we did not measure specific heat for KFe$_2$As$_2$ and RbFe$_2$As$_2$ in present study. Since both of them have even smaller Sommerfeld coefficient than that of CsFe$_2$As$_2$, it is quite reasonable to consider a similar temperature-independent Sommerfeld coefficient for KFe$_2$As$_2$ and RbFe$_2$As$_2$.

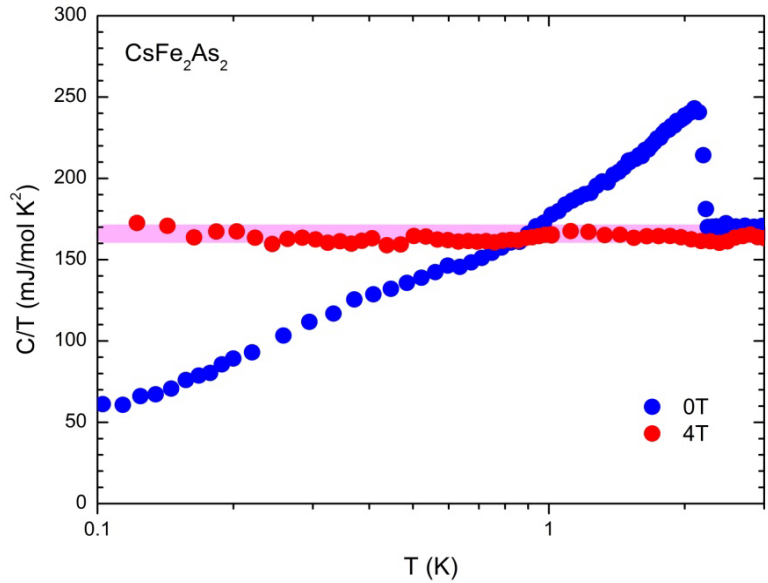

**Fig. S9. Temperature-dependent specific heat measurement for $CsFe_2As_2$.** The external field is applied along c-axis. The specific heat jump due to superconducting transition is completely suppressed under 4 Tesla external field.

**References:**

1. Baek, S. –H. *et al.* $^{75}$As NMR-NQR study in superconducting LiFeAs. *Eur. Phys. J. B* **85**, 159 (2012).

2. Mitrović, V.F. *et al.* Similar glassy features in the $^{139}$La NMR response of pure and disordered $La_{1.88}Sr_{0.12}CuO_4$. *Phys. Rev. B* **78**, 014504 (2008).

3. Yang, Y.-F. *et al.* Scaling the Kondo lattice. *Nature* **454**, 611 (2008).

4. Sarrao, J. L. & Thompson, J. D. Superconductivity in Cerium- and Plutonium-Based '115' Materials. *J. Phys. Soc. Jpn.* **76**, 051013 (2007).

5. Pfleiderer, C. Superconducting phases of f-electron compounds. *Rev. Mod. Phys.* **81**, 1551 (2009).

6. Tafti, F. F. *et al.* Universal V-shaped temperature-pressure phase diagram in the iron-based superconductors $KFe_2As_2$, $RbFe_2As_2$, and $CsFe_2As_2$. *Phys. Rev. B* **91** 054511 (2015).

7. Curro, N. J. *et al.* Scaling in the emergent behavior of heavy-electron materials. *Phys. Rev. B* **70**, 235117 (2004)

8. Stockert, O. *et al.* Superconductivity in Ce- and U-Based "122" Heavy-Fermion Compound. *J. Phys. Soc. Jpn* **81**, 011001 (2012).

9. Kim, J. S. *et al.* Specific heat in KFe$_2$As$_2$ in zero and applied magnetic field. *Phys. Rev. B* **83**, 172502 (2011).

10. Zhang, Z. *et al*. Heat transport in RbFe$_2$As$_2$ single crystals: Evidence for nodal superconducting gap. *Phys. Rev. B* **91**, 024502 (2015).

11. Wang, A. F. *et al*. Calorimetric study of single-crystal CsFe$_2$As$_2$. *Phys. Rev. B* **87**, 214509 (2013).

12. Yang, Y.-F. & Pines, D. Emergent states in heavy-electron materials. *Proc. Natl. Acad. Sci. USA* **45**, E3060–E3066 (2012).